\def\apj{{\em The Astrophysical Journal}}

\def\apjl{{\em The Astrophysical Journal Letters}}

\def\grl{{\em Geophys. Res. Lett.}}
\def\jgr{{\em J. Geophys. Res.}}

\def\mnras{{\em Monthly Notices of the Royal Astronomical Society}}

\def\nat{{\em Nature}}
\def\natphy{{\em Nature Physics}}

\def\nu{{\em Nuclear Fusion}}

\def\prl{{\em Phys. Rev. Lett.}}

\def\pop{{\em Phys. Plasma}}

\def\ssr{{\em Space Sci.~Rev.}}

\documentclass[a4paper]{jpconf}
\usepackage{amsmath}
\usepackage{graphicx}
\usepackage{caption} 
\begin{document}
\title{ A Brief Review on Particle Acceleration in Multi-island Magnetic Reconnection}
\author{H. Che}
\address{University of Maryland, College Park, MD, 20742, USA}
\address{NASA Goddard Space Flight Center, Greenbelt, MD, 20771, USA}
\author{G. P. Zank}
\address{Center for Space Plasma and Aeronomic Research (CSPAR), University of Alabama in Huntsville, Huntsville, AL 35805, USA}
\address{Department of Space Science, University of Alabama in Huntsville, Huntsville, AL 35899, USA}

\begin{abstract}
The basic physics and recent progresses in theoretical and particle-in-cell (PIC) simulation studies of particle acceleration in multi-island magnetic reconnection are briefly reviewed. Particle acceleration in multi-island magnetic reconnection is considered a plausible mechanism for the acceleration of energetic particles in solar flares and the solar wind. Theoretical studies have demonstrated that such a mechanism can produce the observed power-law energy distribution of energetic particles if the particle motion is sufficiently randomized in the reconnection event. However, PIC simulations seem to suggest that the first-order Fermi acceleration mechanism is unable to produce a power-law particle energy distribution function in mildly relativistic multi-island magnetic reconnections. On the other hand, while simulations of highly relativistic reconnections appear to be able to produce a power-law energy spectra, the spectral indices obtained are generally harder than the soft power-law spectra with indices $\sim -5$ commonly observed in the solar wind and solar flare events. In addition, the plasma heating due to kinetic instabilities in 3D magnetic reconnection may ``thermalize" the power-law particles, making it even more difficult for multi-island reconnections to generate a power-law spectrum. We discuss the possible reasons that may lead to these problems. 
\end{abstract}
\section{Introduction}
\vspace*{0.5cm}
Magnetic reconnection (MR) is believed to be an essential process for particle energization in magnetospheric substorms \cite{dungey61prl,baker96jgr,zelenyi10}, solar wind \cite{lin11ssr,khabarova17sg}, and solar flares \cite{priest00book,dennis11ssr,lin11ssr,benz17lrsp}, and is drawing increasing interests in its possible roles in astrophysical phenomena such as the launching of the solar wind \cite{gloeckler03jgr,fisk03jgr}, $\gamma$-ray flares in the Crab Nebula \cite{buhler14,blandford17ssr} or pulsar nebulae in general, and $\gamma$-ray bursts \cite{grb12book,blandford17ssr}.   

\begin{figure}[th]
\includegraphics[scale=0.8,trim=20 430 10 50,clip]{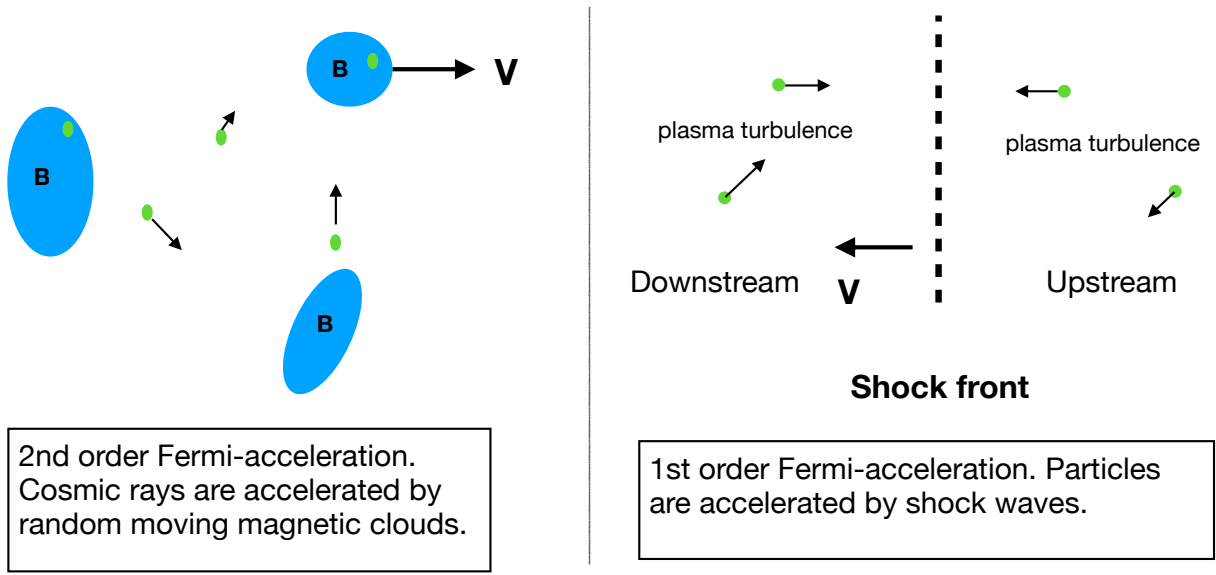}
\caption{Illustration of 2nd and 1st order Fermi-acceleration in astrophysical media and shock waves.}
\label{fermi_acc}
\end{figure}

The mechanisms of particle acceleration in space and astrophysical plasma is still not fully understood \cite{treumann15aapr}.  We need to understand how particles gain energy and how the observed power-law energy distribution $f(W)=W^{-\alpha}$ ($\alpha > 0$) is produced \cite{lin11ssr,oka18ssr}. 
Two types of acceleration mechanisms, namely, the first and second order Fermi acceleration (Fig.~\ref{fermi_acc}), form the foundation of most of the current models. In 1949, Fermi proposed the first acceleration mechanism for the acceleration of cosmic rays \cite{fermi49pr}. In essence fast particles move among wandering magnetic fields in space and spiral around the lines of force until they collide with an irregular moving magnetic cloud and are reflected. On average the particles have an energy gain in each reflection, proportional to the second order of the velocity ratio, i.e., $\delta W= (V/c)^2 W$, where $V$ is the average velocity of magnetic clouds, and hence this mechanism is called the second order Fermi acceleration. The randomness of moving magnetic clouds leads to a power-law energy distribution. The second acceleration mechanism concerns the diffusive shock acceleration \cite{axford77book,bell78mnrasa,bell78mnrasb,blandford78apjl,krymskii77}. The particles are accelerated as they are bounced back and forth between the downstream and upstream of shock waves and energy gain in each crossing is proportional to the first order of $V/c$, where $V$ is the velocity of shock waves, and hence the mechanism is called first order Fermi-acceleration. The particles are scattered during crossings by plasma turbulence generated by the shock, resulting in a power-law energy distribution. 
 
\begin{figure}[th]
\includegraphics[scale=0.8,trim=20 520 10 50,clip]{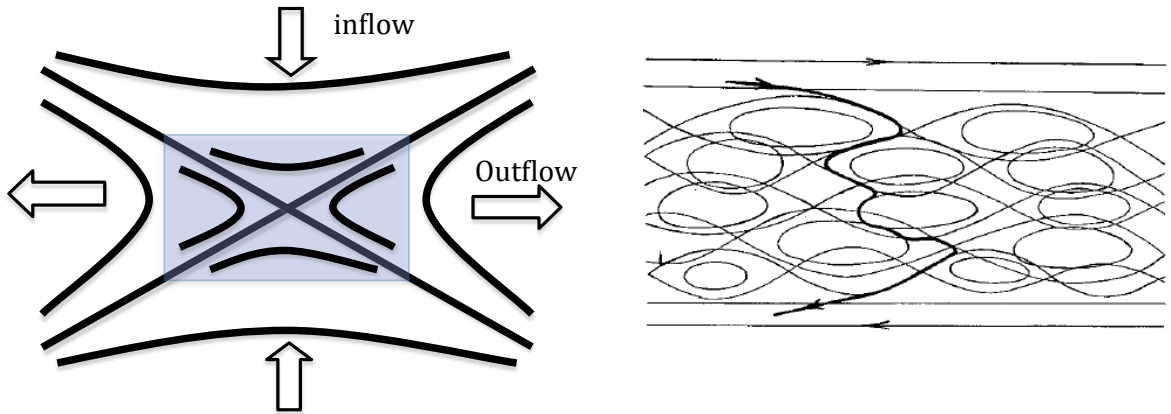}
\caption{Left: Illustration of magnetic reconnection x-line. The shaded blue area is the diffusion region, where charged particles and the magnetic field decouple. Right: Illustration of patchy magnetic reconnection. Solid line: the stochastic wandering of magnetic field lines resulting in particle penetration of magnetic islands (from Galeev, 1986 \cite{galeev86ssr}).  }
\label{reconn}
\end{figure}

Observations of substorms and solar flares show that particles are commonly accelerated  during MR \cite{oieroset02prl,lin11ssr,oka18ssr}, but the acceleration process involved is complex and diverse \cite{biskamp93book,aschwanden02ssr}. In MR the released magnetic energy predominantly transfers to the electromagnetic energy flux, the thermal energy flow,  and the kinetic energy of plasma outflow or the bulk motion \cite{goldman16ssr,le16pop} (Fig.~\ref{reconn}). It is suggested that plasma outflow is accelerated either by the inductive reconnection electric field, such as the perpendicular electric field $\mathbf{V}= -c\mathbf{E}\times \mathbf{B}/B^2$, the parallel electric field \cite{speiser65jgr,biskamp93book,birn09ag,ededal12natphy,kho16grl}, or by the gradient/curvature of magnetic field induced drifts \cite{prit08pop,li17apj,munoz18apj}. The ohmic heating and wave-particle interactions generated by instabilities and turbulence lead to the conversion to thermal energy \cite{che10grl,che11nat,daughton11natphy,moser12nat,eyink13nat,phan16grl}.
The bulk acceleration from MR can not produce a power-law energy distribution. Turbulent heating generated by instabilities does commonly produce a non-Maxwellian velocity distribution function through Landau damping or nonlinear wave-particle interactions, such as phase mixing by electron holes, but a power-law energy distribution is not a common feature \cite{drake05prl,prit08pop,che11pop,che14pop,dahlin14pop,dahlin16pop,dahlin17pop}. To solve this problem, a first-order Fermi-acceleration in multi-island MR was proposed \cite{drake06nat}. This model assumes that particles are repeatedly accelerated by the randomly distributed magnetic islands generated by tearing instability \cite{coppi66prl,kaw79prl}. 

Tearing instability is driven by the magnetic shear in current sheets and can be triggered when the width of current sheet becomes close to ion inertial scale. MR driven by tearing instability is thus called spontaneous MR \cite{galeev86ssr}. Tearing instability produces multiple magnetic islands and the coalescence of islands lead to stochastic wandering of magnetic field lines and percolation of islands \cite{prit79pop,tajima86,schu97pop}, and spontaneous MR is therefore also called patchy MR (Fig.~\ref{reconn}). The stochastic motion of particles caused by the coalescence of islands may result in the power-law energy distribution in principle, which is the essential element in the multi-islands MR acceleration model \cite{drake06nat,zank14apj,leroux18apj}. However, particle-in-cell (PIC) simulations carried by several groups show that similar to shock acceleration, multi-island MR also faces an injection problem. It is  found that for extreme relativistic cases with Lorentz factor $\gamma \gg 1$ and extremely magnetized plasma with $B^2/4\pi mc^2 \gg 1$, a power-law energy distribution function can be obtained \cite{guo14prl,guo16pop,guo16apjl,boro17apj}; but for mildly relativistic cases  with 
$\gamma \sim 1$ and $B^2/4\pi mc^2 \sim 1$, a power-law energy distribution is difficult to achieve, particularly for energetic electrons \cite{dahlin14pop,dahlin16pop,dahlin17pop}, although in some simulations  electrons with lower energy at a few specific regions develop a power-law energy distribution \cite{prit08pop,oka10apj,munoz18apj}.

Since energetic particles accelerated in heliophysics and solar physics are mildly relativistic, the injection problem in multi-island MR acceleration is a serious issue for the mechanism to be applied to particle acceleration in solar flares. This paper discusses progresses and problems in the study of multi-island MR acceleration in the past decade. The basic physics of the multi-island MR acceleration is introduced, the progress and unsolved problems are described, and then we present a summary of the main points and conclusions.

\section{ Particle Acceleration in Multi-island Magnetic reconnection}
\label{basic}
\begin{figure}
\includegraphics[scale=0.7, trim = 50 340 100 70,clip]{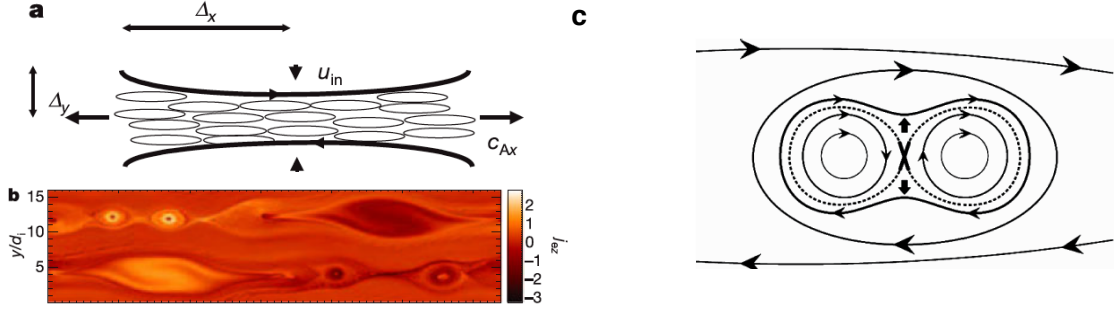} 
\caption{\footnotesize Left Panel: a) Diagram showing volume-filing islands driven by a tearing instability around the magnetic field reversal region. b) PIC simulation of island formation during magnetic reconnection. $j_{ez}$ is the out-of-plane current sheet. (From Drake, Swisdak, Che \& Shay, Nature, 2006 \cite{drake06nat}.) Right Panel: c) a cartoon of a locally oppositely directed magnetic field (lines with arrows) in neighbouring islands experiencing magnetic reconnection (Zank et. al. 2014 \cite{zank14apj}).}
\label{island}
\end{figure}

\vspace*{0.25cm}
The basic picture of multi-island MR is schematically shown in Fig.~\ref{island}. The size of the islands varies from electron to ion-scales. Opposite directed magnetic fields between neighbouring islands develop localized MR. These small scale MR events can be triggered by turbulence due to the twists and braids of magnetic flux tubes. In large-scale MRs such as those associated with solar flares, plasmoids generated by current-driven instabilities such as the tearing instability fill the magnetic field reversal region. In multi-island MR, beside the acceleration produced by reconnection induced electric field, the contraction and merging of the islands also lead to the acceleration of particles \cite{drake06nat,oka10apj,bian13prl,zank14apj,zank15apj,leroux15apj,leroux16apj}. 
\subsection{Acceleration via Contraction of Magnetic Islands in Magnetic Reconnection}
\begin{figure}
\includegraphics[scale=0.6, trim = 30 300 50 70,clip]{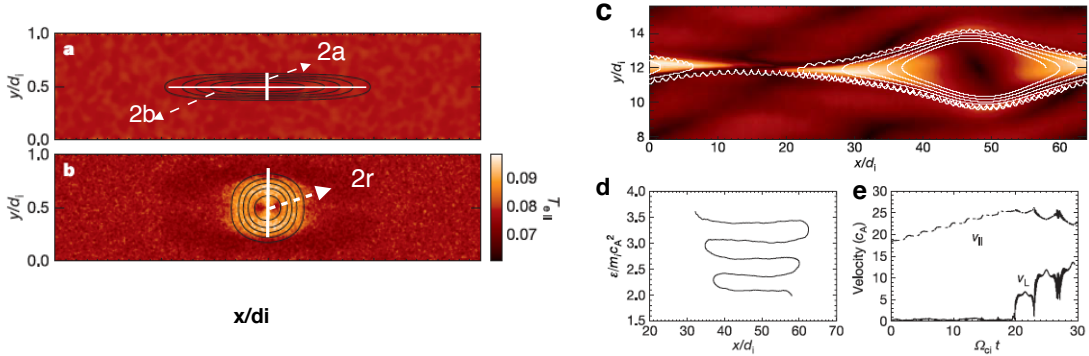} 
\caption{\footnotesize Left Panel: a 2D simulation of a squashed magnetic flux bubble. Right Panel: Test particle orbit and the sudden energy gain of electrons at the ends of island due to the curvature drift.  (From Drake, Swisdak, Che \& Shay, Nature, 2006 \cite{drake06nat}). }
\label{contract}
\end{figure}

\vspace*{0.25cm}
To illustrate how island contraction accelerates electrons, let's consider a 2D anti-parallel MR in the $xy$ plane. Without external forces, an elongated magnetic island will gradually become round due to its internal magnetic tension.  This effect is illustrated in a 2D simulation in Fig.~\ref{contract} (a, b). Initially the magnetic island is a very narrow ellipse, and later it becomes a circle with radius $r$. In this process, the magnetic flux $\psi$ is approximately conserved such that $B_{x,a} a \approx B_{x,b}r$, where $B_{x,a}$ and $B_{x,b}$ are the magnetic field as shown in panels (a) and (b), respectively. In an incompressible plasma, the volume of the magnetic island is conserved, so that $\pi ab=\pi r^2$.  The perimeter of magnetic island decreases during this process. The perimeter  $l$ of an eclipse approximately is $l \approx \pi (a+b) (1+ 3h /(10 + \sqrt{4-3h})) > \pi (a+b) \ge 2\pi \sqrt{ab} = 2\pi r $, thus the perimeter decrease as the island evolves from eclipse to round --- a reason we call the contraction of island.

The loss of magnetic energy  $\bigtriangleup\epsilon$ can be estimated by \cite{zank14apj}:
\begin{equation}
\bigtriangleup\epsilon=\frac{b-a}{b}\epsilon_a,
\end{equation}
where $\epsilon_a=\pi a b B_{x,a}^2$ is the initial magnetic energy stored in the island.  

In the simulation in Fig.~\ref{contract} (a, b), the temperature of the trapped electrons has increased as the island contracts, indicating the magnetic energy is converted to the kinetic and thermal energy of particles.  It should be noted that the volume of the magnetic island is generally not conserved and magnetic energy can accumulate during contraction. During the contraction, electrons are accelerated in directions both parallel and perpendicular to the magnetic field. The detailed derivation of the acceleration process can be found in \cite{drake06nat} and \cite{zank14apj}. Here we describe the basic picture.  

The trapped electrons circulate along the magnetic field lines of the contracting islands and gain energy in each circle (Fig.~\ref{contract} (c, d) ). Since the island contraction is relatively slow,  the second adiabatic invariant $J = m_e \oint v_{\parallel}ds \sim m_e v_{\parallel} l$ implies
\begin{equation}
\frac{dv_{\parallel}}{dt} =-v_{\parallel}\frac{1}{l}\frac{dl}{dt}.
\label{inv2}
\end{equation}
Here $v_{\parallel}$ is the parallel velocity of a single electron, $ds$ is the displacement along magnetic field lines, and $l$ is the perimeter of the magnetic island. Since $dl/dt<0$ due to contraction, $dv_{\parallel}/dt>0$, and the trapped electrons are accelerated as they bounce inside the island. 

The first adiabatic invariant $\mu = m_e v_{\perp}^2/2 B=constant$ and the conservation of magnetic flux $\psi\sim Bl$ describe the electron acceleration in the perpendicular direction:
\begin{equation}
\frac{dv_{\perp}^2}{dt}=\frac{v_{\perp}^2}{B}\frac{dB}{dt}, \qquad  \frac{dB}{dt}=-B\frac{dl}{dt}.
\label{inv1}
\end{equation}
Again since $dl/dt<0$,  $dv_{\perp}^2/dt>0$. This energy gain process is shown in a test particle simulation (Fig.~\ref{contract} (d, e)). In panel (d) we see that the electron energy jumps when the electron arrives at each endpoint of the magnetic island. The perpendicular velocity, i.e. the gyration velocity abruptly increases as $v_{\parallel}$ slightly decreases, implying that the energy transfer is from the parallel to perpendicular.

Assume the velocity of contraction of an island with perimeter $l$ is $V_l= dl/dt$, both Eq.~(\ref{inv1}) \& (\ref{inv2}) can be rewritten in the form of  
\begin{equation}
\frac{dW}{dt}\propto \frac{V_l}{c} \frac{1}{\tau_c}W,
\end{equation}
where $W=m_e v^2/2$ and $\tau_c = l/c$. Thus the contraction of island is in essence a 1st order Fermi acceleration. 

The changing of curvature of magnetic islands can accelerate particles in the perpendicular direction, especially, at the ends of islands (Fig.\ref{contract}), as the particles trapped within magnetic islands and slowly spiral along magnetic field lines, i.e. $v_{\parallel}\gg v_{\perp}$. The curvature drift is 
\begin{equation}
\mathbf{v}_{c}=\frac{mv_{\parallel}^2}{eB} \mathbf{b}\times (\mathbf{b}\cdot\nabla \mathbf{b}),
\end{equation}
where $\mathbf{b}=\mathbf{B}/B$. The curvature drift contributes to the perpendicular velocity. Once $v_{\perp}$ is sufficiently large, the particle can escape the trap of magnetic island and migrate to another magnetic island.

\subsection{Acceleration via Merging of Magnetic Islands}
\vspace*{0.25cm}
The merging of islands produces localized MR (Fig.~\ref{island}). For simplicity, we first neglect the reconnection electric field so that the first and second adiabatic invariants hold. Following Fermo et al. \cite{fermo10pop}, let's assume two round islands with initial areas $A_1$ and $A_2$,  which are merged into a single island with area $A_f=A_1+A_2$.  The circumference of the merged island satisfies $l=(l_1^2+l_2^2)^{1/2}<l_1+l_2$, which implies that the $dl/dt<0$ in Eq.~(\ref{inv2}). Consequently, the merging leads to the electron acceleration in the parallel direction.

During merging, the magnetic flux from the two islands is not additive but is equal to the larger of the two initial fluxes, max$(\psi_1,\psi_2)$ \cite{fermo10pop,zank14apj}. Before merging, the total magnetic energy of the two islands is $W_i=B_1^2 \pi l_1^2+B_2^2\pi l_2^2=\pi( \psi_1^2+\psi_2^2)$. The magnetic energy of the merged island $W_f=\pi \max(\psi_1, \psi_2)^2<W_i$. Thus merging of islands is an alternative path to efficiently dissipate magnetic energy and accelerate electrons in the parallel direction. If we consider the localized electric field caused by the reconnections, part of the magnetic energy is transferred to the kinetic energy of electrons via electric field acceleration.

\subsection{ First-order Fermi Acceleration and Power-Law Energy Distribution}
\vspace*{0.25cm}
The MR electric field accelerate particles via $d\mathbf{v}/dt= q/m \delta \mathbf{E}$. In the contraction and merging  of islands, first and second adiabatic invariants and magnetic flux are approximately conserved and causes parallel and perpendicular acceleration.
To obtain a power-law particle energy distribution, particles must experience sufficient stochastic motion. If the contraction of multi-islands acceleration dominates, particles are scattered by randomly distributed magnetic island or turbulence, then the acceleration is a first-order Fermi-acceleration. Zank et\thinspace al. studied how various acceleration mechanisms in multi-island MR driven by tearing instability contribute to the index of power-law energy distribution.
\begin{figure}
\includegraphics[scale=0.63, trim = 70 400 20 55,clip]{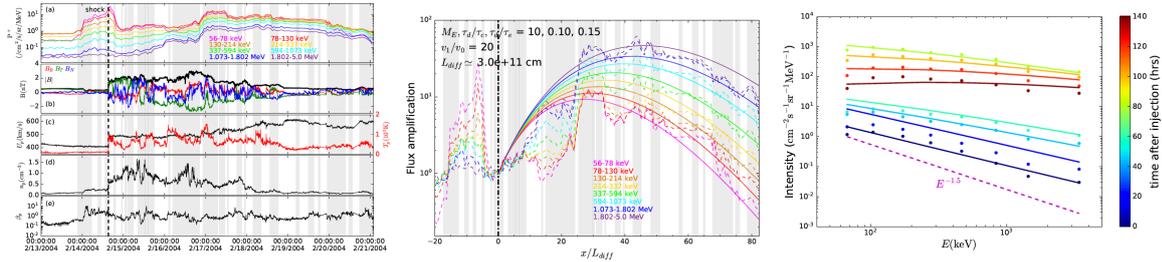} 
\caption{{\bf Left} Ulysses observations. The vertical gray lines show the regions where we have observed (using a Grad-Sharafnov reconstruction technique) the presence of magnetic islands.  {\bf Middle:} The comparison of theoretical (dashed) and observational (solid) particle flux. {\bf Right:} The comparison of theoretical (solid) and observational (dotted) differential intensity spectrum. (Zhao, Zank et\thinspace al. 2018 \cite{zhao18apj}) }
\label{zfig}
\end{figure}

By expanding the 6D Liouville equation, Zank et al. \cite{zank14apj} obtain:
\begin{equation}
 \frac{\partial f}{\partial t}+\mathbf{v}\cdot\frac{\partial f}{ \partial \mathbf{x}}+\frac{\mathbf{F}_L}{m}\cdot\nabla_v f 
 = \frac{\delta f}{\delta t}\bigl)_s-\frac{(\mathbf{F}-\mathbf{F}_L)}{m}\cdot\nabla_v f- f\nabla_v \cdot  (\frac{\mathbf{F}}{m}), 
\label{zank1}
\end{equation}
where $\mathbf{F} \equiv \mathbf{F_L}+\mathbf{F_c}+\mathbf{F_m}+\mathbf{F_E} $,  $\mathbf{F_L}\ = \mathbf{v}\times \mathbf{B}/c $ is the Lorentz force, $\mathbf{F_c}$ is the force caused by the contraction of islands, $\mathbf{F_m}$ is the force caused by the merging of islands, and $\mathbf{F_E}$ is the force caused by localized reconnection inductive electric field $\delta \mathbf{E}$. While the Lorentz force is a function of $v$, it can be shown that $\nabla_v (\mathbf{v}\times \mathbf{B})=0$. On the other hand, $\nabla_v\cdot \mathbf{F}_c/m=2\eta_c$,  $\nabla_v\cdot \mathbf{F}_m/m=\eta_m-\eta_m = 0$. The pitch-angle scattering term is assumed to have the simplest possible form $\delta f/\delta t)_s=\partial ((1/\tau_s)(1-\mu^2)(\partial f/\partial \mu))/\partial \mu$, where $\mu=\cos\theta$ is the cosine of the particle pitch angle. $\tau_s\sim \Omega=eB/m$ is the characteristic time-scale of pitch angle scattering.

Zank et{\thinspace}al. \cite{zank14apj} considered the MR acceleration in a super-Alfv\'enic flow, e.g., the solar wind. In this case, the convective flow velocity term typically is much larger than the corresponding advective Alfv\'enic velocity, allowing the latter term to be safely neglected in the transport formulation. To eliminate the electric field  $\mathbf{E}=-\mathbf{U}\times \mathbf{B}$ due to the flow bulk motion, we rewrite the equation in the plasma flow frame using $\mathbf{v} \equiv \mathbf{c}+\mathbf{U}$. 
Averaging Eq.~(\ref{zank1}) over the gyro-angle, expanding the gyrophase-averaged distribution function $f$ by Legendre polynomials, then the Laplace transformation yields a general solution for a nearly isotropic particle distribution in the steady state of Eq.~(\ref{zank1}):


%
\begin{align}
\nonumber
f(x, c/c_0) =   \frac{n_0}{8\pi c_0^2 v_E}&(\frac{c}{c_0})^{-(3+M_E+2\tau_{diff}/(3\tau_c)M_E^2)/2}
 \exp(-\frac{2\tau_{diff}}{3\tau_c}M_E\frac{x}{L_{diff}}) I_0(\Phi) \times & \\
 & H[\ln (\frac{c}{c_0})] \times H [\ln(\frac{c}{c_0})+\frac{2}{M_E}\frac{x}{L_{diff}}],\label{zank4}  
 \end{align}    

\begin{equation*} 
\Phi  =  [\frac{\tau_{diff}}{3\tau_c}  M_E^2(M_A-3+\frac{\tau_{diff}}{3\tau_c}M_E^2)]^{1/2}\times [(\ln(c/c_0)^2+\frac{4}{M_E}\ln(c/c_0)\frac{x}{L_{diff}}]^{1/2},
 \end{equation*}
where $\tau_c\equiv 1/\eta_c$ is the contraction time, $\tau_{diff}=\kappa/U^2$ is the diffusion time, $L_{diff}=\kappa/U$ is the diffusion length, $M_E=U/v_E$ and $v_E$ is the reconnection electric field accelerated velocity; $I_0$ is the modified Bessel function of order 0 and $H(x)$ is the Heaviside step function. We can see that the velocity power-law index depends on $M_E$ and $\tau_{diff}/\tau_c$. $M_E$ is a unique quantity in MR due to the formation of particle beams. Eq.~(\ref{zank4}) has successfully explained the enhancement of energetic particle flux produced at the downstream of a shock at 5 AU by multi-island model \cite{zhao18apj} (Fig.\ref{zfig}).   If we consider only magnetic island contraction,  the steady-state solution yield a velocity power-law with index $\alpha = 3[1+ \tau_c/(8\tau_{diff})]$~\cite{zank14apj}, implying the longer diffusion time produces a harder power-law.

\section{ Outstanding Issues in the Current Studies of Multi-island Magnetic Reconnection}
\label{prog}
\vspace*{0.25cm}

\begin{figure}
\includegraphics[scale=0.68, trim = 60 530 30 55,clip]{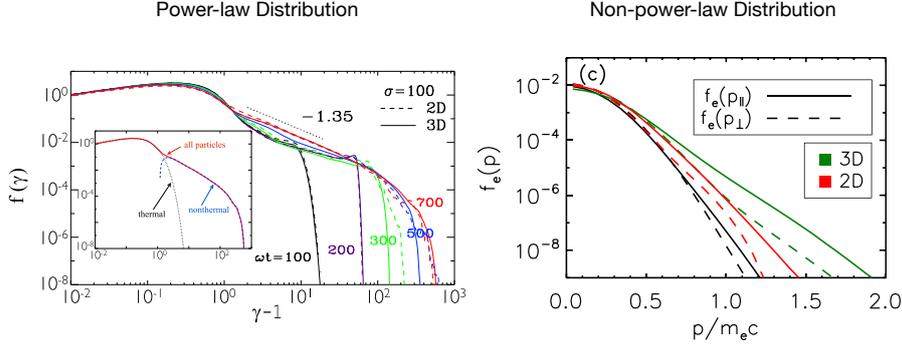} 
\caption{{\bf Left} Power-law electron energy distribution from highly relativistic PIC simulation; Guo et al.\cite{guo14prl}. {\bf Right:} Non-power law electron energy distribution from mildly relativistic PIC simulation; Dahlin et al.\cite{dahlin17pop}. }
\label{plaw}
\end{figure}

Analytical studies \cite{drake13apjl,zank14apj,zank15apj,leroux15apj,leroux16apj,leroux18apj} have shown that a power-law particle energy distribution can be achieved in multi-island MR if the motion of particles can be well randomized. However, in PIC simulations how to produce a power-law particle energy distribution remains a main issue, especially, how to reproduce the observed power-law spectra found in both solar flares and solar wind. PIC simulations  \cite{sironi14apjl,guo14prl,li15apjl,guo16apjl,werner16apjl} that have successfully produced power-law energy spectra have assumed extreme relativistic conditions with typical Lorentz factors $\gamma \sim 100$ to $10^3$, and magnetic field $\sim 10^{12}$~G (the corresponding ratio of electron gyro-frequency and plasma frequency is $\Omega_{ce}/\omega_{pe} \sim 100$). These simulations may not be relevant to the bulk of energetic electrons in the solar corona and solar wind whose velocities are typically $0.2c-0.5c$ ($\gamma \sim 1.02-1.3$) and magnetization $\Omega_{ce}/\omega_{pe}<1$.  PIC simulations with similar system sizes but essentially non-relativistic were so far unable to generate the power-law energy spectrum \cite{dahlin14pop,dahlin16pop,dahlin17pop}. As an example, we show the electron energy distribution functions obtained from PIC simulations carried by two independent groups in Fig.~\ref{plaw}.

It is unclear why mildly relativistic PIC simulations fail to produce the power-law energy distribution.  One possible reason is that it is easier for highly relativistic particles to escape from the trapping of a single magnetic island and hence the particles can interact with many different islands. The repetitive acceleration of the particles in different islands makes the motion of particles more stochastic. In mildly relativistic PIC simulations, particles are confined in the magnetic islands for a relatively long period of time and particle transport among magnetic islands is slow. For example, in the PIC simulation  by Drake et al.\cite{drake06nat}, the average time for an electron to be confined in the magnetic island is about 30 $\Omega_i^{-1}$ (see Fig.~\ref{contract}, right panel). For a simulation domain of about 200~$d_i$, the maximum length of the simulation free from complications caused by boundary condition is about 200 $\Omega^{-1}$. The mean number of islands an electron can interact with during the whole simulation is only about 7, which may be too small to randomize the electron population. 

\begin{figure}
\includegraphics[scale=0.66, trim = 55 275 80 80,clip]{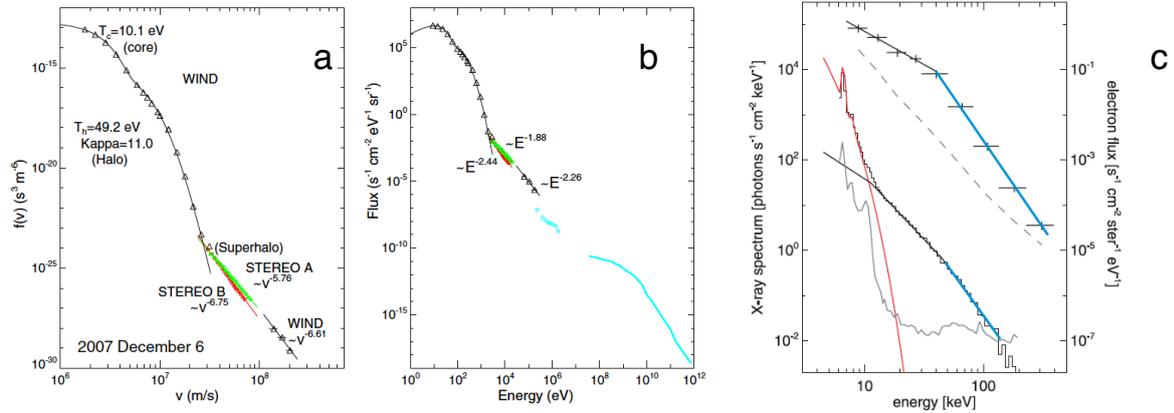} 
\caption{\footnotesize \textbf{a)} The quiet-time electron velocity distribution function f(v) at 1AU;  \textbf{b)} the flux spectrum. a) and b) are obtained from the solar wind observation of Wang et. al.\cite{wang12apjl}. The electron energy ranges from $\sim$ 9 eV to $\sim$ 200 eV as measured by \textit{WIND} (black) and \textit{STEREO} (green) A (green) and B (red) on 2006 December 6. The solid lines represent the fit by a sum of a Maxwellian and a Kappa distribution to the solar wind core and halo distribution and a power-law fit to the superhalo. In panel (b), the blue diamonds/curves represent the electron flux spectra of the interplanetary electrons at 0.2-2 MeV and galactic cosmic ray at $\sim $ 30 MeV-1 TeV. Panel \textbf{c)} is a broken power-law energy spectrum fit to the data (blue line) from the paper by S. Krucker, 2007\cite{krucker07apjl}.  }
\label{superhalo}
\vspace{10pt}
\end{figure}

Another problem with current PIC simulations is that none can produce the observed steep power-law above the spectral break in flare energetic electrons (Fig.~\ref{superhalo}) \cite{holman03apjl,krucker07apjl,krucker09apj,lin11ssr,james17mnras}. The typical energy spectral index obtained in the aforementioned relativistic simulations is $\sim -1.3$, which is close to the observed index below the spectral break ($\sim 50-100$~keV), but far from the index $-3.9\pm 0.9$ above the break. This index can not match the observed energy spectrum of superhalo electrons in the solar wind with a index $\sim -3$. According to the steady-state solution from Zank et al. \cite{zank14apj}, the hard power-law is likely a result of long diffusion time.  The softer power-law spectra in solar flares and solar wind indicates particle transport is likely very fast.

Most of the aforementioned simulations are conducted in 2D. In 3D MR, it is almost unavoidable that various kinetic instabilities are triggered. This results in plasma heating that may suppress the generation of a power-law energy distribution function \cite{oka10apj,dahlin16pop,dahlin17pop,munoz17pop}. How to reconcile this issue with the observed power-law spectra is probably the key to understand particle acceleration in solar flares and solar wind.   

\section{Summary and Conclusion}
\vspace*{0.25cm}
Contraction of multi-island during MR is an emerging mechanism for particle acceleration in space and astrophysics \cite{drake06nat,lin11ssr,zank14apj}. In the past decade extensive analytical and numerical studies have been carried out. The followings are the main points of this paper:
\begin{itemize}

\item The acceleration from contraction of magnetic island during MR is proportional to the first-order of $V_l/c$, where $V_l$ is the contraction velocity of the island \cite{drake06nat}.

\item Particles attain a power-law energy distribution $f(W)\propto W^{-\alpha}$ through repetitive interactions with multiple islands. Therefore, this mechanism is a first-order Fermi-acceleration. Analytical studies show that the contribution from contraction to the index of power-law $\alpha$ is related to the ratio of contraction time and particle diffusion time $\tau_c/\tau_{diff}$, implying longer diffusion time leads to a harder power-law \cite{zank14apj}.

\item PIC simulations can produce a power-law in highly relativistic MR events \cite{guo14prl}, but fail to produce a power-law energy distribution in mildly relativistic  MR,  which is more relevant to particle acceleration in solar wind and solar flares \cite{dahlin17pop}.

\item In addition, the power-law energy distribution function obtained by highly relativistic PIC simulations is so far too hard ($\alpha \sim 1-2$) to explain the energetic particle energy distribution observed in the solar wind ($\alpha \sim 5$) and solar flares.

\item The apparent contradiction between analytical calculations and the PIC simulations implies that the assumption of the particle diffusion time in multi-island MR is long, which may explain the hard power-law in relativistic MR acceleration simulations, but in mildly relativistic MR accelerations in solar flares and solar wind, the diffusion time is likely to be much shorter. What can cause such short diffusion time is the subject of future studies. 
\end{itemize}

\section{Acknowledgement}
HC is partly supported by NASA Grant No.NNX17AI19G. The simulations and analysis were supported by the NASA High-End Computing (HEC) Program through the NASA Advanced Supercomputing (NAS) Division at Ames Research Center.
\section*{References}
\vspace*{0.25cm}
\providecommand{\newblock}{}

\end{document}